\documentstyle[     
preprint,      
aps]{revtex}      
\begin{document}      
\draft

\title{Frenkel and charge transfer excitons in C$_{60}$     
}      
      
\author{Martin Knupfer and J\"org Fink}      
\address{      
Institut f\"ur Festk\"orper- und Werkstofforschung Dresden, D-01171      
Dresden, Germany      
}

\date{\today}      
\maketitle      
\date{\today}      
      
\begin{abstract}       
We have studied the low energy electronic excitations of C$_{60}$      
using momentum dependent electron energy-loss spectroscopy in transmission.      
The momentum dependent intensity of the gap excitation      
allows the first direct experimental determination of the energy of the      
$^1H_g$ excitation and thus also of the total width of the multiplet      
resulting from the gap transition. In addition, we could elucidate  
the nature of the      
following excitations - as either Frenkel or charge transfer excitons.     
\end{abstract}      
      

C$_{60}$ has attracted a lot of      
attention during the last years due to its remarkable      
physical properties such as superconductivity      
\cite{rev7,rev14}, non-linear optical      
properties      
\cite{koopmans93,kuhnke98}, or        
magnetism \cite{rev12}.      
Despite the large number of studies devoted to      
C$_{60}$ there remain unanswered question      
regarding a complete microscopic understanding of      
its properties. Since it also acts as a model      
substance for the description of $\pi$-electron      
systems in general and their interaction in solids and interfaces, 
its properties are also of general      
value with respect to the entire class of molecular      
materials with $\pi$-governed electronic structures.      
     
\par     
One of the questions which has remained unanswered up to now  
is connected with the      
electronic excitations across the energy gap in C$_{60}$ at about 1.8 to 2 eV.      
For the icosahedral point group      
the gap transition of C$_{60}$ splits into a multiplet with 4 components      
having $^1T_{1g}$, $^1T_{2g}$, $^1G_{g}$ and $^1H_{g}$ symmetry \cite{negri88}.      
While there exist several studies of the low energy components and their      
excitation energies \cite{hartmann95,muccini96a},      
the energy of the highest lying $^1H_{g}$ excitation and thus the      
total width of the multiplet have not been experimentally determined yet,      
although it has important consequences for the understanding of      
optical properties such as e.g. second or third harmonic generation (SHG, THG) \cite{eder96,bech99}.      
Additionally, it also provides an important test quantity regarding the applicability of      
different theoretical approaches to describe these electronic excitations.      
Theoretical predictions for the multiplet width      
spanning a wide range from 150 meV to 600 meV have been      
reported \cite{eder96,negri92,shirley96}.      
Furthermore, the nature of the energetically higher lying excitations that are observed      
as absorption maxima at about 2.4 eV and 2.7 eV is still under discussion.      
From a number of experiments both have been controversially assigned to either      
Frenkel excitons (on-ball) or charge transfer excitons      
\cite{eder96,dick94,petelenz94,kazaoui95,wei97,kazaoui98,munn98}.      
    
\par     
In this study we present a detailed analysis of the momentum dependent      
electronic excitations of C$_{60}$ as  
obtained using electron energy-loss spectroscopy (EELS)      
in transmission. We show that the intensity variation of the low lying     
excitation can be used to derive the excitation energy and the mean     
extension of the $^1H_{g}$ excitation. This additionally gives the first experimental     
determination of the multiplet width arising from the gap transition.     
Furthermore, the nature of the two excitations at about 2.4 and 2.7 eV could     
be clarified.

\par     
For our EELS studies C$_{60}$ films with a thickness of      
about 1000 \AA\ have been prepared by evaporation      
onto KBr single crystals under high vacuum conditions. After      
deposition, the films are floated off in destilled water, mounted      
on standard electron microscopy grids and transferred into the      
spectrometer. Prior to the measurements the films were characterized      
using electron diffraction showing that they were polycrystalline.      
The electron diffraction and EELS measurements were performed in transmission at      
room temperature using a 170 keV spectrometer described elsewhere.\cite{jfeels}      
The energy and      
momentum resolutions were chosen to be 110 meV and 0.05 \AA$^{-1}$.      
The raw data have      
been corrected for contributions from the elastic line.      
In order to obtain the optical conductivity from the measured loss functions we have      
performed a Kramers-Kronig analysis.      
Our results for low momentum transfers (optical limit)     
are in very good agreement with those obtained from optical      
measurements \cite{ren91,pichler92}.

\par     
The response function in EELS in transmission is proportional      
to the dynamic structure factor $S({\bf q},\omega)$, i.e. EELS can      
provide information on the spatial structure of the electronic excitation      
under consideration.       
The matrix element, $M$, for EELS is proportional to                  
$<1 \mid exp(iqr) \mid 0>$ which can be expanded to                  
                  
\begin{equation}                   
M \propto \sum_{n}\frac{i^n}{n!} (q<r>)^n <1 \mid (\frac{r}{<r>})^n \mid 0>.   \label{G2}                   
\end{equation}                  
                  
Hereby, the introduction of a mean radius $<r>$ allows one to      
separate the characteristic dimensionless $(q<r>)$-dependence      
of the matrix element from the (now also dimensionless) constant excitation      
probability $<1 \mid (r$/$<r>)^n \mid 0>$. In the case of excitations      
with a {\it specific} multipole character (e.g. dipole excitations) the latter has      
a finite value only for the corresponding $n$ (e.g. $n = 1$), i.e. one     
can derive the mean radius of the excitations from their moment dependence.       
The mean radius $<r>$ gives a measure for the extension       
of the electron-hole wave function $\Psi_{eh}({\bf r})$ in the               
excited state which represents the probability amplitude to find the electron at         
a certain distance $<r>$ assuming that the hole is fixed.         
Thus, the momentum dependence of the excitation intensity $I_n$ ($\propto$ $|M|^2$) of                    
an excitation with a {\it specific} multipole character can be written as     
                  
\begin{equation}                   
I_n \propto \frac{n!^{-2}(q<r>)^{2n}}{N}, \hspace{0.8cm} N = \sum_{n}\frac{(q<r>)^{2n}}{n!^2}.                      
\label{G3}                   
\end{equation}                  
                   
$N$ is a normalization factor which guarantees the oscillator strength sum rule.

In Fig. 1 we show the intensities $I_n$ as a        
function of $q<r>$ for $n$ = 1 (dipole excitations) and      
$n$ = 2 (quadrupole excitations).      
Fig. 1 demonstrates that from the intensity variation of an excitation      
with, for example, quadrupole character as a function of momentum transfer, $q$,      
one can derive the mean radius of the excited electron-hole      
pair (exciton).      
     
\par    
In Fig. 2 the optical conductivity,      
$\sigma$, of C$_{60}$ is shown between 0.6 and 3.2 eV   
for various momentum transfers.      
The optical conductivity has been derived performing a   
Kramers-Kronig analysis of the loss function which was   
published previously    
\cite{romberg93}.      
We now demonstrate how these results can be used to determine the      
multiplet width of the gap excitation and the nature of the two following      
excitations. In Fig. 2 strong intensity variations  
with increasing momentum transfer are visible.       
At low momentum transfers the spectrum consists of a small shoulder     
at about 2.1 eV and two further structures located at about 2.45 and     
2.8 eV. While the 2.1 eV shoulder develops into a clear peak with increasing     
momentum transfer the intensity of the two higher lying excitation decreases.     
At a momentum transfer of about 0.8 \AA$^{-1}$ the feature     
at 2.1 eV reaches its maximal intensity before it starts to decrease again,     
whereas the other two structures show the opposite behavior.

\par        
In order to obtain the momentum dependent intensity        
variation of the electronic excitations as                   
observed in Fig. 2 quantitatively, we have modelled the optical        
conductivity with a sum of Lorentz oscillators:                  
                  
\begin{equation}                   
\sigma(\omega) = \epsilon_0\sum_{j}\frac{\omega^2f_j\gamma_j}                   
{(\omega_j^2-\omega^2)^2+\omega^2\gamma_j^2},   \label{G1}                   
\end{equation}                   
                  
with $\omega_j$ being the energy position, $\gamma_j$ the width        
and $f_j$ the oscillator strength of the                   
corresponding excitation. 
The energy positions were fitted to obtain a best agreement within the 
entire $q$-dependent series and kept the same for all $q$. The width 
of the excitations was also kept constant for all $q$ and models the finite 
lifetime as well as phonon satellites of the electronic excitations. 
The result of this fit for the        
momentum dependent intensity of                   
the low energy features seen in Fig. 2 is shown in Fig. 3.      
     
\par    
Fig. 3a depicts the intensity variation of the gap      
transition at 2.1 eV. As expected, it clearly shows the behavior of a      
dipole forbidden excitation.     
The EELS response function in the gap region is almost solely caused     
by the $^1H_{g}$ excitation because    
it can be reached via an      
electric quadrupole transition while the other multiplet components can      
only be reached via magnetic dipole or even higher order transitions \cite{koopmans93}.      
Consequently, a comparison of the intensity variation of the gap transition      
shown in Fig. 3a and Fig. 1 gives a direct measure of the mean radius      
of the $^1H_{g}$ excitation of $<r>$ $\sim$ 2.8 \AA.      
This value is somewhat smaller than one would expect for the mean      
distance $\bar{d}$ of two particles moving independently on a sphere with the      
radius, $R$, of a C$_{60}$ molecule which is $\bar{d}$ = $(4R_{C_{60}}/\pi$) $\sim$      
4.5 \AA. This indicates that the electron-hole pair is considerably      
excitonic in agreement with other results and predictions \cite{lof92}.      
From the mean radius one can also derive an estimate for the      
exciton binding energy, $E_B$:      
     
\begin{equation}                
E_B \sim \frac{1}{4\pi\epsilon_0\epsilon_r}\;\frac{e^2}{<r>}\;,                 
\end{equation}                
      
which is screened by the static dielectric constant        
$\epsilon_r$ ($\sim$ 4 for C$_{60}$  \cite{ren91,pichler92}).             
This simple consideration leads to a                 
binding energy $E_B$ of about 1.3 eV also in good agreement with      
other results \cite{lof92}. We note that local, unconventional screening effects, which  
have been discussed for finite size systems \cite{shirley96,brink97}, play a minor role for the  
binding energy derived above as the bare and the effective Coulomb repulsion  
are very similar at a distance of 2.8 \AA\, \cite{brink97}.  
     
Moreover, the fact that the $^1H_{g}$ excitation dominates the      
EELS response function in the region of the energy gap yields a first direct experimental access to its      
excitation energy which can be derived from the first maximum      
in Fig. 2 ($q$ around 0.8 \AA$^{-1}$) to be 2.1 eV.      
Together with optical studies of the other multiplet components \cite{muccini96a}      
we can determine the total width of the gap excitation multiplet of      
C$_{60}$ to be about 260 meV.     
Thus, the multiplet width of the gap transition of C$_{60}$     
is significantly smaller than predictions from calculations \cite{negri92,shirley96}     
which indicates that, independent of the exact approach,   
the models used to describe the electronic excitations      
of C$_{60}$ (and other $\pi$ electron systems) tend to     
overestimate electron interaction effects.     
     
In Fig. 3b and 3c we show the $q$-dependent  
intensity changes of the excitations visible at     
about 2.45 and 2.8 eV. The excitation occurring at 2.45 eV does not show     
any significant momentum dependence which suggests that it is     
not of a pure multipole but of mixed character. Since intra-molecular     
excitations in C$_{60}$ can all be classified as either {\it gerade} or     
{\it ungerade} we conclude that the electronic excitation appearing    
at 2.45 eV in C$_{60}$ is a charge transfer excitation resulting in the final state     
hole and electron sitting on different molecules.     
In contrast, the spectral weight of the excitation at 2.8 eV decreases     
with increasing momentum transfer which is consistent with a     
dipole allowed excitation. We therefore attribute the 2.8 eV     
feature in the optical conductivity to an intra-molecular or     
Frenkel exciton. The second intensity maximum visible in   
Fig. 3c at about 1.3 \AA$^{-1}$ is probably caused by   
a resonance effect in the solid. Provided the mean radius of  
an excitation is commensurate with the lattice spacing further  
intensity maxima can occur at $q \sim 2\pi/<r>$.

\par    
Our assignment of the two excitations at 2.45 eV and 2.8 eV to     
charge transfer and Frenkel excitons, respectively, is in agreement with     
a recent comprehensive analysis of the excited states of  C$_{60}$      
using optical absorption and luminescence, electroabsorption and      
photoconductivity \cite{kazaoui98}.     
It is also qualitatively in line with theoretical calculations \cite{shirley96}     
which predicted the charge transfer excited states to occur about 150 meV     
above the $^1H_{g}$ excitation.

\par     
To summarize, analyzing the momentum dependent intensity     
variation of the low lying electronic excitations of C$_{60}$     
we could determine the total multiplet width of the gap transition to     
be about 260 meV, which is smaller than predictions from     
calculations. Additionally, the mean radius ($\sim$ 2.8 \AA) and the binding energy of     
the $^1H_{g}$ exciton ($\sim$ 1.3 eV) could be derived.     
The nature of the next higher lying excitations could also be elucidated;     
the excitation at 2.45 eV gives a     
charge transfer exciton while at 2.8 eV a     
Frenkel exciton is formed.

\par

\par      
\vspace{1cm}      
\par      
{\bf Acknowledgement}      
\par      
We are grateful to M. S. Golden and G. A. Sawatzky for fruitful discusssions and  
to H. Romberg and E. Sohmen for help during the measurements.    
\par

\newpage      
 \parindent0cm

\begin{figure}      
\caption{Intensity of a pure dipole or  quadrupole excitation in EELS      
as a function of the reduced parameter $q<r>$ (see text).     
}      
\end{figure}      
      
\begin{figure}      
\caption{Low energy optical conductivity $\sigma$ of C$_{60}$ as      
a function of momentum transfer $q$ in steps of 0.1 \AA$^{-1}$.      
The curves are offset in y-direction.     
}      
\end{figure}      
      
 \begin{figure}      
\caption{The experimentally determined      
intensity variation of (a) the gap excitation and (b,c) the following      
electronic excitations of C$_{60}$ as a function of momentum transfer $q$.     
The excitation energies are given in the corresponding panel. 
The solid line in panel (a) represents the theoretical expectation for a 
quadrupole excitation with a mean radius of 2.8 \AA\ (see also Fig. 1).    
}      
\end{figure}

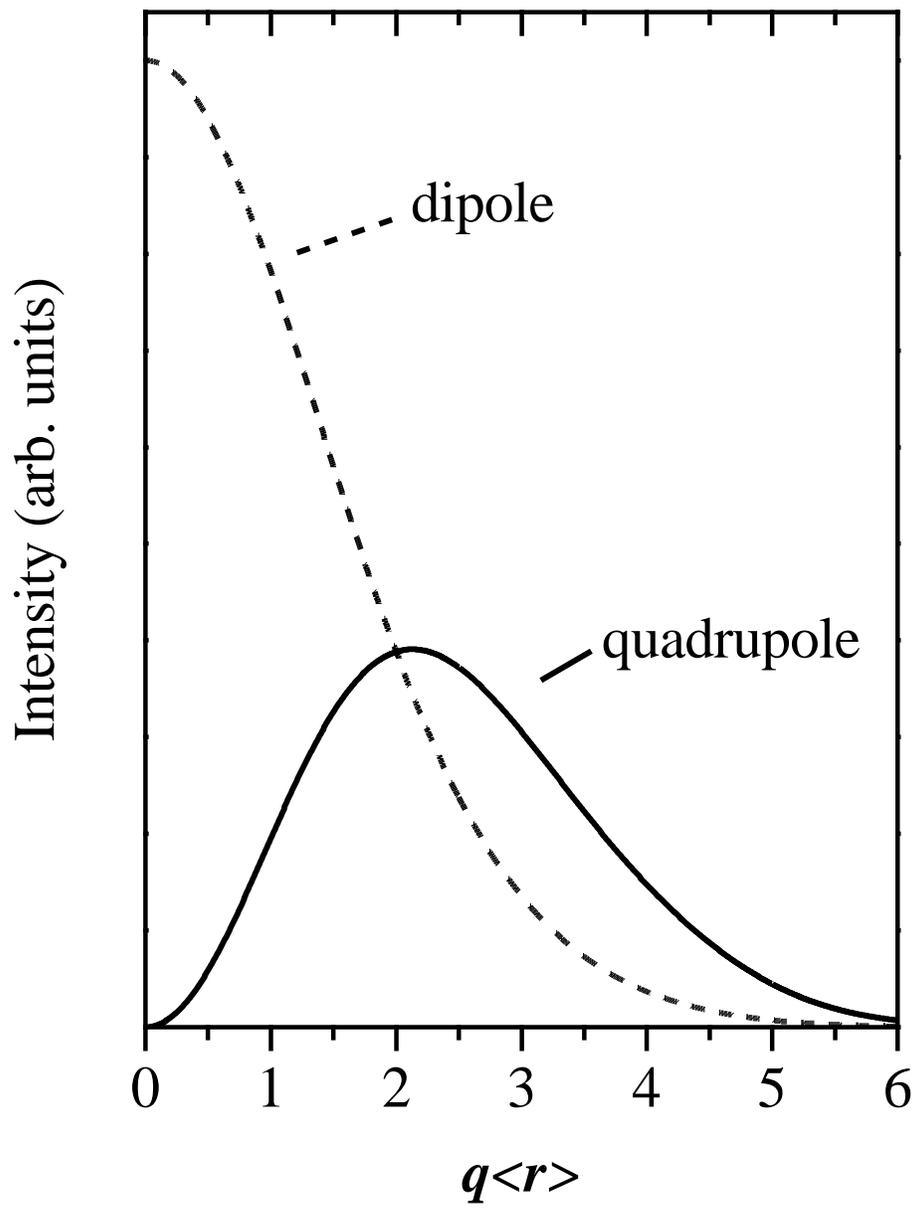
\end{document}